# Substrate Dependent Synergistic Many-Body Effects in Atomically Thin Two Dimensional WS$_2$

*Shreyasi Das*[1]*, Rup K. Chowdhury*[2,3#]*, Debjani Karmakar*[4]*, Soumen Das*[5]*, Samit K. Ray*[2]

[1]School of Nano Science and Technology, Indian Institute of Technology Kharagpur, Kharagpur-721302, India

[2]Department of Physics, Indian Institute of Technology Kharagpur, Kharagpur-721302, India

[3]Department of Ultrafast Optics and Nanophotonics, IPCMS, CNRS, Strasbourg-67034, France

[4]Technical Physics & Prototype Engineering Division, Bhabha Atomic Research Center, Mumbai 400085, India

[5]School of Medical Science and Technology, Indian Institute of Technology Kharagpur, Kharagpur-721302, India

## Abstract

Mott transition has been realized in atomically thin monolayer (ML) of two dimensional semiconductors (WS$_2$) via optically excited carriers above a critical carrier density through many body interactions. The above nonlinear optical transition occurs when excited electron hole pairs in ML WS$_2$ continuum heavily interact with each other followed by transformation into a collective electron hole plasma phase (EHP), by losing their identity as individual quasiparticles. This is manifested by the alluring red-shift-blue-shift crossover (RBC) phenomena of the excitonic peaks in the emission spectra, resulting from the synergistic attraction-repulsion processes at the Mott-transition point. A systematic investigation of many-body effects is reported on ML $WS_2$, while considering the modulated dielectric screening of three different substrates, viz., silicon dioxide, sapphire, and gold. Substrate doping effects on ML $WS_2$ are discussed using the Raman fingerprints and PL spectral weight, which are further corroborated using theoretical DFT calculations. Further the substrate dependent excitonic Bohr radius of ML $WS_2$ is extracted via modelling the emission energy shift with Lennard-Jones potential. The variation of Mott point as well as excitonic Bohr radius is explained via substrate induced dielectric screening effect for both the dielectric substrates, which is however absent in ML $WS_2$ on Au. Our study therefore reveals diverse many-body ramifications in 2D semiconductors and offers decisive outlooks on selecting the impeccable substrate materials for innovative device engineering.

## I.    INTRODUCTION

Excitonic complex induced many-body phenomena control the optical and electrical properties of 2D semiconducting transition metal dichalcogenides (TMDs), making them inimitable candidates for nanoelectronics [1,2], nanophotonics [3,4], optoelectronics [5–7], and valleytronics [8–10] devices. The optical properties of TMDs are strongly influenced by the dielectric environment and thus are dependent on the choice of substrates, which vary widely depending upon the end applications [11–13]. The Coulomb potential inside photo-generated excitonic quasiparticles and many-body phenomena are strongly affected via environmental dielectric permittivity, as the electric line of forces extend beyond the boundaries of atomically thin 2D TMD layers [14,15]. This substrate induced Coulomb-screening effects lead to band-gap renormalization and binding energy reduction of excitons in 2D monolayers, whereas substrate induced unintentional carrier doping plays a significant role on the emission spectral weight, providing a wide design flexibility of 2D-materials based optical devices [16–19]. A comparative study with suspended $MoS_2$ flakes has addressed the substrate induced doping effects on photoluminescence (PL) emission [18], while the effect of dielectric screening on the emission characteristics of $WS_2$ monolayer on different substrates [20] could be explained with band renormalization phenomena. However, a detailed investigation on the substrate induced many-body effects at high carrier density regime is still lacking on 2D TMDs.

The modulation of strain [1], carrier density [21,22], dielectric environment [23,24] and variation of temperature [25,26] play important roles to alter the many-body effects in a 2D system, giving rise to the appearance of higher-order bound states (e.g. trions), line-shift of quasiparticle resonance, and anomalous behavior in the emission spectral weight of quasiparticles [26]. With increasing electron-hole densities from photogenerated carrier doping, Coulombic potential between charges get efficiently screened due to the exchange and correlation effect, leading to band-gap renormalization and the reduction of binding energy of quasiparticles [27–29]. On the other hand, in the high carrier density regime, excitons start to interact with each other and lose their identity to create a new collective phase called electron-hole-plasma (EHP) [30]. This transition from an insulating gas of excitons at lower densities to a metal like EHP state at higher carrier densities, is attributed as Mott transition. At a critical density ($n_M$), where Mott transition takes place, the number of excitons become so high that the average distance between them decreases to their exciton Bohr radius ($r_B$) [31] limit. Further, in this critical density regime, the conductivity of the material undergoes percolation driven metal-insulator transition, which can be realized via temperature dependent transport behavior in the 2D layered systems [32–34]. The coexistence of different many-body effects at an elevated carrier concentration can lead to an anomalous red-shift-blue-shift crossover (RBC) of excitonic resonance, which was reported earlier in case of optical doping via transient absorption spectroscopy [35,36]. though their steady-state emission characteristics are yet to be addressed. Due to the reduced dimensionality and relatively longer lifetime (~picosecond) of excitonic complexes, a higher carrier density is achievable in 2D TMDs even with a continuous wave laser beam excitation in steady-state [37,38]. Moreover, the effect of various dielectric environment on 2D TMDs at the Mott transition point opens up new possibilities for device fabrications [39].

In this work, we report a comparative study on many-body quasiparticle dynamics for 2D monolayer (ML) $WS_2$ flakes on two different dielectric ($SiO_2$ and $Al_2O_3$) and a metallic (Au) substrate through emission and Raman spectroscopy techniques. We established a correlation between (i) carrier density induced, and (ii) substrate dependent many-body effects on $MLWS_2$. The anomalous behavior in PL spectral weight variation and resonance peak-shift associated with the increase of optical doping (n-type) and dielectric screening are discussed for all three samples. The RBC of the excitonic peaks is clearly observed for high carrier density regime (~2-3 $\times$ $10^{12}$ $cm^{-2}$) at the Mott-transition point for the insulating ($SiO_2$/Si and $Al_2O_3$) substrates, which is absent for Au substrate. The origin of these RBC phenomena is further elucidated using band renormalization due to interplay between two distinct many-body effects: (i) Coulomb screening, and (ii) Pauli Blocking. Analogous to the hydrogen atom model, RBC is explained beyond the Mott density regime from the attraction-repulsion crossover of exciton-exciton interactions in 2D $WS_2$, which provides an estimation of the excitonic Bohr radius of $WS_2$ with varying dielectric environments.

# I.  EXPERIMENTAL METHODS: SYNTHESIS AND CHARACTERIZATION

Monolayer $WS_2$ flakes were obtained by mechanical exfoliation from a commercial $WS_2$ crystal (2D Semiconductors, Inc. Scottsdale, AZ, USA) using Scotch magic tape (3M Inc. USA) and directly exfoliated onto three substrates from the tape.as schematically described in Fig. S1. Pieces of commercially purchased 300 nm thick $SiO_2$ on highly doped Si wafer ($1\times1$ cm$^2$) were cleaned with acetone, propanol and DI water in an ultrasonic bath followed by $N_2$ gun drying. The commercially available mirror polished $Al_2O_3$ substrates were also cleaned by acetone and DI water in an ultrasonic bath to remove dust and contaminants. On the other hand, the Au coated $SiO_2$/Si substrate was fabricated by PVD technique with e-beam evaporation of Au using a HIND HIVAC system. A 10 nm thick Au layer was grown on clean $SiO_2$ substrate prior to exfoliation. Direct exfoliation of $WS_2$ on different substrates was employed intentionally to avoid any extra doping inside the material from polymer assisted dry or wet transfer techniques. An optical microscope (DM 2500M with a 100X, 0.9 numerical aperture (NA) objective, Leica) equipped with a color camera was used to locate and identify the monolayer $WS_2$ sheets with the naked eye. Optical contrast based identification of $MLWS_2$ was performed using an optical imaging in reflection geometry combined with an image analysis software (ImageJ) [40,41]. The optical contrast profiles are generated using ImageJ in the R-channel grey scale image, which confirms the formation of $MLWS_2$ flakes on different substrates. Further an atomic force microscopy (AFM) (Agilent Technology, Model: 5500) was used to image the $WS_2$ sheets in tapping mode in air to confirm the layer number by measuring the thickness of the nanosheets. The corresponding AFM topological images and line scan height profiles of the $WS_2$ flakes reconfirm the monolayer exfoliation, with the thickness of the flake being $\sim 0.8$ nm [42]. Detailed investigations of substrate induced effects on material characteristics were carried out by Raman and photoluminescence (PL) spectroscopy techniques, using a fiber coupled confocal micro-Raman/PL Spectrometer (WITec alpha-300R, Inc. Ulm, Germany) with a continuous wave 532 nm laser excitation source. Room temperature power-dependent PL study was performed using an 100x objective (0.8 N.A.) with spot size ~ 0.5µm. The power of the laser was calibrated using an optical power meter from Newport (Model no: 1916-R) (MKS Instruments, Inc. (NASDAQ: MKSI)).

## II.   RESULTS AND DISCUSSIONS

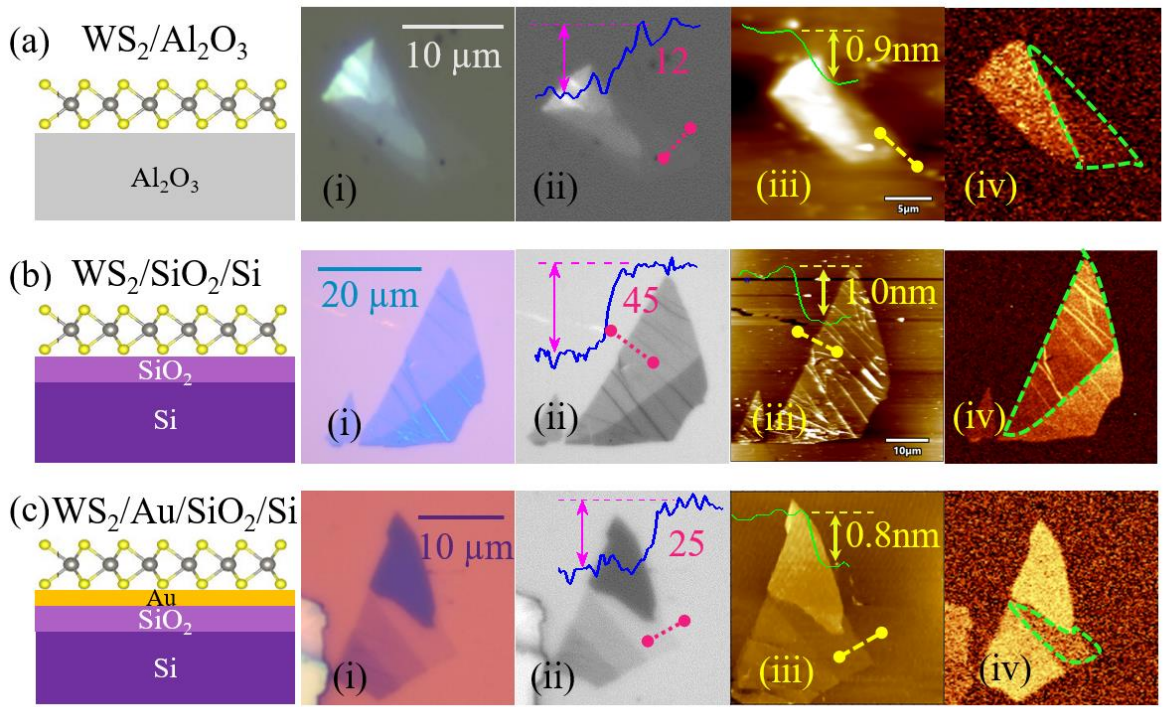

FIG.1. (i) Optical microscopic image, (ii) R-channel grey scale image, (iii) AFM topographic image and (iv) Raman mapping image ($E_{2g}^1$ peak) of corresponding monolayer $WS_2$ flakes on three substrates: (a) $Al_2O_3$ substrate, (b) $SiO_2/Si$ substrate and (c) $Au/SiO_2/Si$ substrate. In the inset of grey scale image (ii), contrast profile has been shown along the pink dashed line extracting data using Image-J software and the height profile of the ML region shown in the inset of AFM

topographic image (iii) along the yellow dashed line. The green dashed areas in Raman mapping images (iv) represent the ML region on each substrate.

ML WS$_2$ flakes were mechanically exfoliated from a commercially available bulk crystal using scotch magic tape (3M) and directly transferred on freshly cleaned substrates: (a) crystalline sapphire (Al$_2$O$_3$), (b) dielectric SiO$_2$/Si (oxide thickness ~ 285 nm) and (c) conducting Au coated (~ 10 nm) SiO$_2$/Si substrate, as schematically shown in Figs. 1(a-c) show optical micrographs (i), R-channel gray scale images (ii) with contrast profiles and AFM topography images along with height profiles (iii), which confirm direct exfoliation of ML WS$_2$ on each substrate. Further the Raman mapping of in-plane E$_{2g}^1$(M) mode (~ 351 cm$^{-1}$) of WS$_2$ flakes on all three substrates are illustrated in Figs. 1(a-c) (iv), where the green dashed areas are the ML region with reduced Raman intensity compared to their bulk counterparts [43]. A comparative Raman study for all samples is presented in Fig. 2(a) with the intense in-plane E$_{2g}^1$(M) Raman peaks at ~ 351 cm$^{-1}$ and the out-of-plane vibrational peaks A$_{1g}$ at ~ 421 cm$^{-1}$ [44]. With the variation of substrate, a clear shift of the out-of-plane (A$_{1g}$) vibrational Raman modes is recorded, as shown in Fig. 2(c). The softening of A$_{1g}$ peak in Raman spectrum with substrate variation has a direct correlation with the charge transfer mechanism in 2D TMDs [45]. The increased electron doping leads to the filling-up of anti-bonding states of the conduction band (CB), mostly made up with d$_{z}^2$ orbitals of transition metal atoms, making the bonds weaker. Thus, A$_{1g}$ mode, which preserves the symmetry of the lattice, softens and reveals an increasing n-type doping in WS$_2$. The A$_{1g}$ peak softens by 5.069 cm$^{-1}$ from Au to SiO$_2$/Si substrate, showing the increase in n-type doping in ML WS$_2$, as corroborated by the band structure calculation [Fig. 3] (Detailed discussion in APPENDIX A).

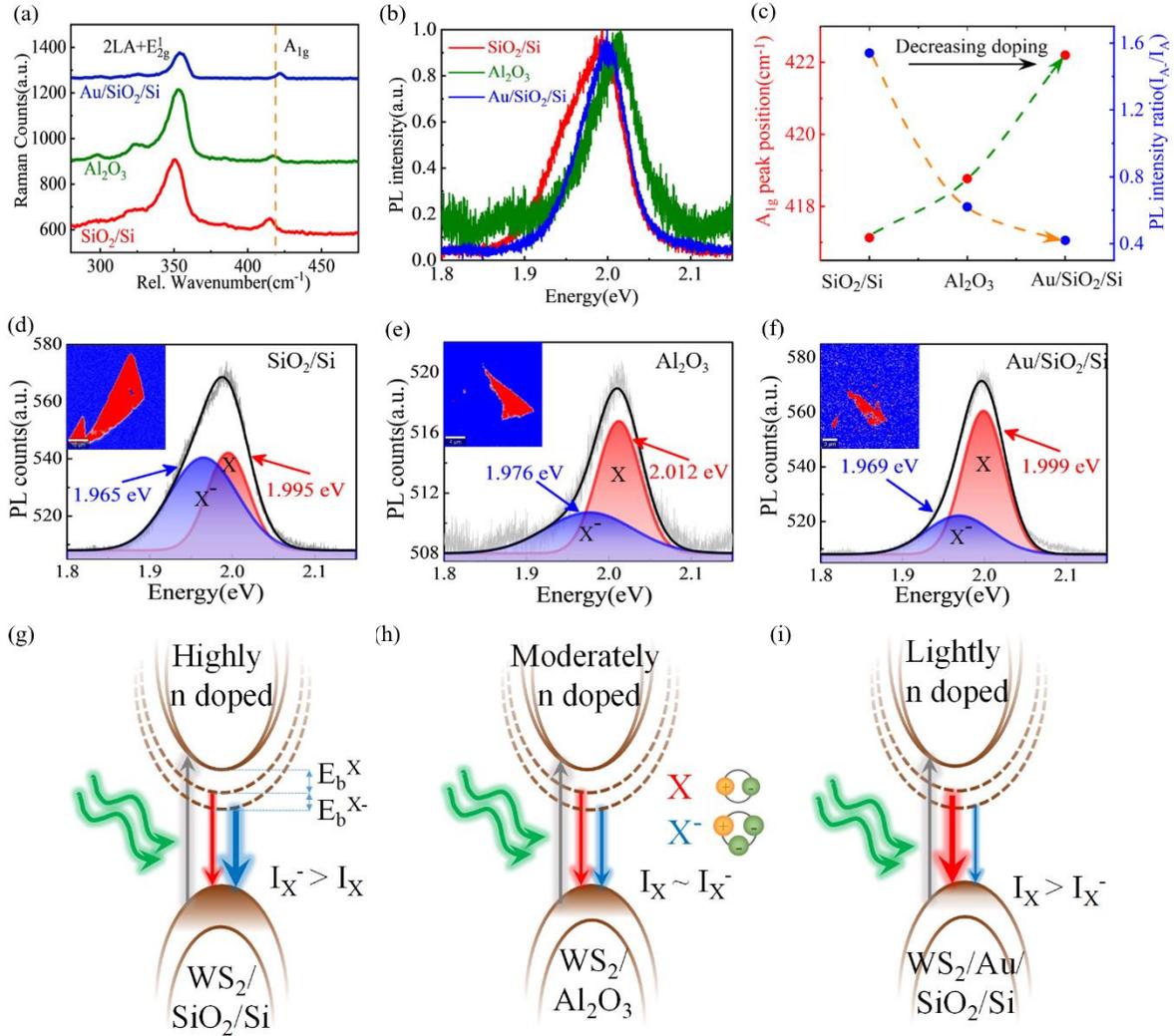

FIG.2. (a) Comparative Raman spectra of ML WS$_2$ on all three substrates, collected at a very low laser power excitation (0.3 MW/cm$^2$). (b) Comparative normalized PL emission spectra at a low laser power density (0.3 MW/cm$^2$) of ML WS$_2$ flakes on three substrates; SiO$_2$/Si (Red line), Al$_2$O$_3$ (Green line) and Au/SiO$_2$/Si (Blue line) showing clear peak-shift and variation of FWHM. (c) The variation of A$_{1g}$ Raman peak and PL intensity ratio of trion-to-exciton for different substrates, revealing substrate induced doping. The green and orange dashed lines are drawn for the guidance of the eye. (d)-(f) Deconvoluted photoluminescence spectra of WS$_2$ on three different substrates

showing two distinct peaks corresponding to excitonic (red line) and trionic (blue line) emissions at a low laser power density ($\sim$0.3 MW/cm$^2$). The black line symbolizes the cumulative fit of two distinct peaks. The PL mapping of excitonic peak originating from the monolayer region is illustrated in the top left inset. (g)-(i) Schematic representation of substrate doping induced exciton-to-trion intensity variation for SiO$_2$/Si, Al$_2$O$_3$ and Au/SiO$_2$/Si substrates, respectively.

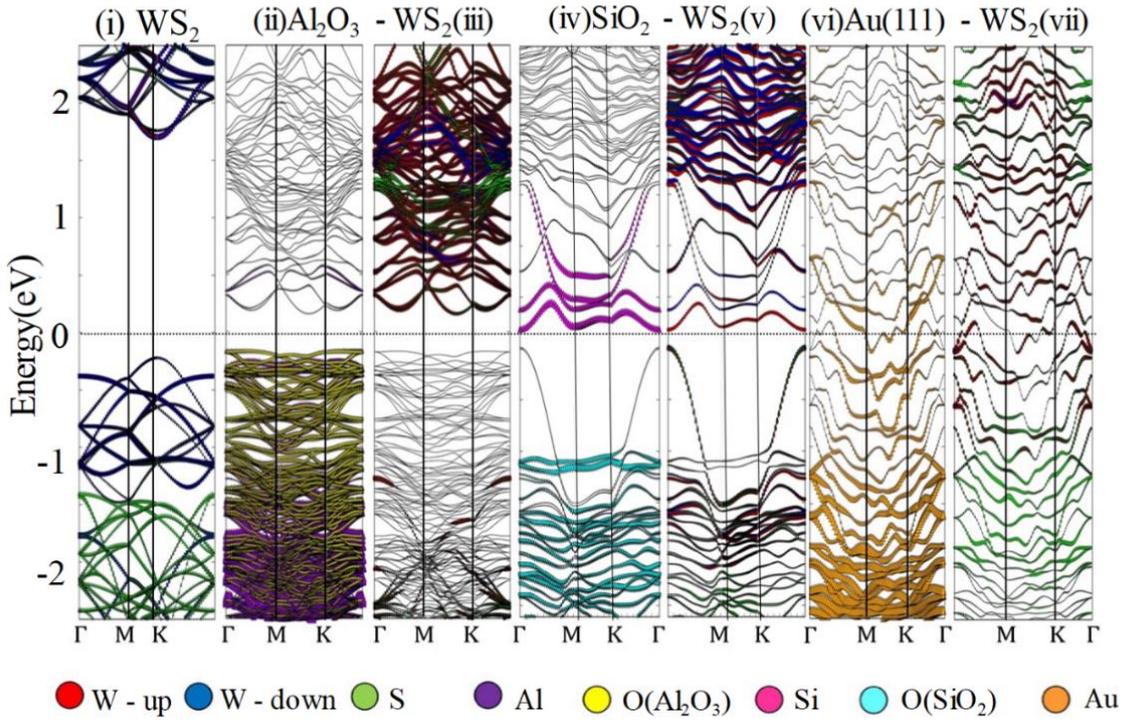

FIG.3. Orbital-projected band structures (i) for WS$_2$ ML with W-up (up-spin channel), W-down (down-spin channel) $5d$ orbitals and S-$3p$ orbitals, (ii) Al-$3s$ and $3p$ and O-$2p$ projected bands for Al$_2$O$_3$/WS$_2$, (iii) W-up (up-spin channel), W-down (down-spin channel) $5d$ and S-$3p$ projected bands for Al$_2$O$_3$/WS$_2$, (iv) Si-$3s$ and $3p$ and O-$2p$ projected bands for SiO$_2$/WS$_2$, (v) W-up (up-spin channel), W-down (down-spin channel) $5d$ and S-$3p$ projected bands for SiO$_2$/WS$_2$, (vi) Au-

$5d$ and $6s$ projected bands for Au[111]/WS$_2$, (vii) W-up (up-spin channel), W-down (down-spin channel) $5d$ and S-$3p$ projected bands for Au[111]/WS$_2$.

Moreover, the PL spectral weight of ML WS$_2$ on various substrates may also differ because of the substrate induced charge doping into the material. Comparing the normalized PL spectra of ML WS$_2$ for all the samples [Fig. 2(b)], a broader emission peak with a full-width-half-maximum (FWHM) of 83.5 meV is perceived in case of WS$_2$ on SiO$_2$/Si. To evaluate the substrate induced variation of photoluminescence emission energy and spectral weight, we deconvoluted all the spectra into two Gaussian peaks to acquire optimal line-shape of the excitonic quasiparticles for a fixed laser power density (P$_d$) ~ 0.3 MW/cm$^2$, as depicted in Figs. 2(d-f). The PL spectrum for WS$_2$/SiO$_2$/Si (k-SiO$_2$ ~ 3.9) [46] shows A-excitonic emission at ~ 1.995 eV, corresponding to direct band-to-band transitions of ML WS$_2$ at K (and/or K′) point of the Brillouin zone, and trionic-A$^-$-emission at ~ 1.965 eV with a binding energy of ~ 30 meV, which are in good agreement with previously reported results of ML WS$_2$ [47,48]. However, for a higher-k dielectric substrate like Al$_2$O$_3$ (k ~ 9) [46], the A and A$^-$-peaks are found at 2.012 eV and 1.976 eV, respectively, which indicates that the PL peak energy of all the quasiparticles gets blue-shifted with increasing k-value of the substrates. This blue shift of the resonance energy is a consequence of the dielectric screening induced band renormalization and binding energy reduction (detailed discussion in APPENDIX B). The PL mapping images of A-peaks [see top left inset of Figs. 2(d-f)] reveal strong emissions from the ML regions of WS$_2$ on three different substrates. Further, the differential reflectivity plot of WS$_2$ on three substrates confirms the A-excitonic emission, as depicted in Fig. S2. However, trion peaks are not realized from the room temperature reflectivity spectra in agreement with previous reports [14,41,49]. Moreover, in our room temperature (300 K) PL and reflectivity measurements, the absence of B-excitonic peak (~ 2.4 eV) has a direct consequence of large valence band splitting due to spin-orbit coupling in ML WS$_2$. At 300 K, most of the carriers occupy the highest valence band maximum (VBM) level containing higher density of states, resulting in the absence of B-excitonic PL even at a higher laser power [50]. The sublinear variation in the double logarithmic plot of deconvoluted PL peak integrated intensity with laser power [Fig. S3] confirms the neutral excitonic (A) and trionic (A$^-$) emission for all three substrates. It is important to emphasize, that for all three substrates the integrated intensity variation of each peak is fitted with the power law equation only in the low power density regime to avoid photodoping induced many-body effects. The extracted integrated intensity of quasiparticles from each deconvoluted spectrum is tabulated in Table I, from which we observe trion-to-exciton integrated intensity ratio for SiO$_2$/Si, Al$_2$O$_3$, and Au/SiO$_2$/Si substrates to be 1.54, 0.62, and 0.42, respectively. Further, we have calculated the carrier concentration ($n_e$) in ML WS$_2$ from integrated PL intensity ratio of trion and exciton using the law of mass action [42],

$$n_e = \left(\frac{I_{A^-}}{I_A}\right)\left(\frac{\gamma_A}{\gamma_{A^-}}\right)\left(\frac{4m_A m_e}{\pi \hbar^2 m_{A^-}}\right)(k_B T) exp\left(\frac{E_b^{A^-}}{k_B T}\right) \qquad (1)$$

Where, $\hbar$ is the reduced Planck's constant, $k_B$ is the Boltzmann constant, $E_b^{A^-}$ is the trion binding energy, $m_A(m_{A^-})$ are the effective masses of exciton (trion), and $\gamma_A(\gamma_{A^-})$ are the radiative decay rates of excitons (trions). Here, $m_{A^-} = 2m_e + m_h$ and $m_A = m_e + m_h$, where $m_e$ and $m_h$ are the effective masses of electrons and holes in WS$_2$, respectively. The corresponding areal carrier concentration ($n_e$) for above three substrates are found to be ~ 3.62 ×10$^{13}$, 1.54×10$^{13}$, and 0.98×10$^{13}$ cm$^{-2}$ in ML WS$_2$, which contribute to the fraction of trion formation and spectral weight variation with changing substrate materials. As represented schematically in Fig. 2(g), for SiO$_2$/Si substrate, the maximum carrier concentration (denoted by brown shade in the valence band) leads to the formation of higher number of trions (blue arrow) than exciton (red arrow) and $WS_2$ behaves as highly n-doped. In case of $Al_2O_3$ substrate, $WS_2$ can be considered as moderately n-doped with the formation of almost equal quantity of trions and excitons [Fig. 2(h)]. On the other hand, due to draining of charges to the Au film, $WS_2$ on Au/SiO$_2$/Si substrate behaves as lightly n-doped with the exciton density greater than that of trion, as depicted in Fig. 2(i). The existence of charged trions along with excitonic features is a consequence of the negative charge transfer due to unintentional substrate induced doping in ML $WS_2$ for all samples at a low laser power density [51]. In general, electrons move towards a material with lower Fermi level from that having a higher Fermi level, and the excess electrons inside the system interact with neutral excitons to form charged excitons or trions. The formation of trion reduces the exciton density in the system, leading to a decrease in PL emission efficiency with the broadening of peak width. Therefore, a lower integrated intensity ratio between trion ($I_A^-$) and exciton ($I_A$) signifies lower n-doping level in monolayer $WS_2$ and the resultant PL efficiency is eventually dominated by neutral excitonic emission. As a result, $SiO_2$ substrate induces maximum n-type doping with highest trion-to-exciton intensity ratio ($\sim 1.54$) along with a greater FWHM as compared to other two substrates. Here, it may be noted that our theoretical DFT calculations also suggested [Fig. 3] higher doping from $SiO_2$ substrate than that of $Al_2O_3$, which is corroborated with the experimental findings. However, Au coated $SiO_2$ substrate reveals maximum n-type doping in our DFT studies. This can be explained via metallic nature of the interface between Au and $WS_2$, as the photo-generated excess electrons inside ML $WS_2$ are transferred to the Au coated substrate through the interface [Fig. 3]. As a consequence, it hinders the trion formation processes in the 2D $WS_2$, leading to the lowest trion-to-exciton intensity ratio of 0.42 as well as the reduced FWHM of the PL spectrum, as observed experimentally in our case.

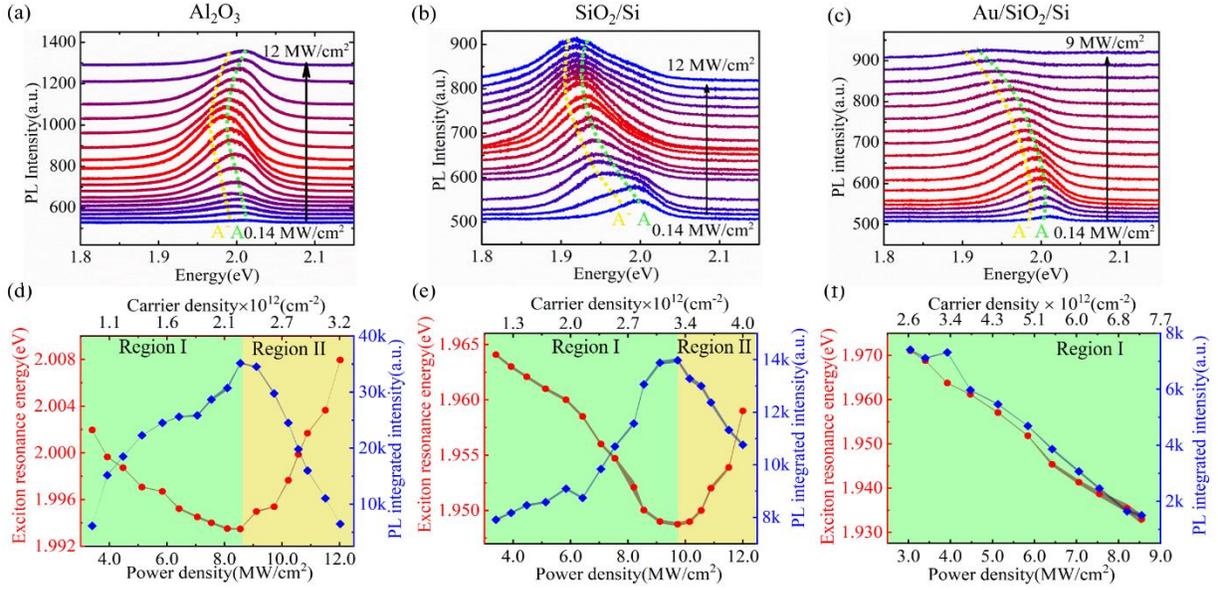

FIG.4. PL spectra of the ML WS$_2$ for different excitation powers on (a) Al$_2$O$_3$, (b) SiO$_2$/Si, and (c) Au/SiO$_2$/Si substrates. The colour gradient signifies the intensity variation, i.e. first increases up to a maximum value (from blue to red) and then decreases further (from red to blue). The green (yellow) dotted line is a guide to the eye of exciton (trion) peak energy variation with laser power density. The variation of exciton resonance energy and integrated intensity (plotted for higher excitation power regime) with excitation induced charge density in ML WS$_2$ on (d) Al$_2$O$_3$, (e) SiO$_2$/Si and (f) Au/SiO$_2$/Si substrates. The red-shift-blue-shift crossover and integrated intensity fall of exciton peak at Mott transition point are indicated for dielectric substrates Al$_2$O$_3$ and SiO$_2$/Si via interface between two regions, Region I (green) and Region II (yellow).

For room temperature realization of Mott insulator to metal transition with investigation of many body effects on the nature of excitonic quasiparticles and their substrate dependency, the intensity variation of PL emission and resonance peak shift have been investigated for each substrate with elevated carrier population by increasing the laser power density. This has a direct consequence on the many-body effects of ML $WS_2$. Here, it is important to mention that, we purposefully carried out all the power dependent PL measurements with an accumulation time of 0.5 sec and 10 accumulations, to avoid prolonged laser exposure induced photodoping and local heating effects during the experiment. Moreover, we have also ensured to collect individual PL spectrum at different locations on the ML $WS_2$ flakes for each laser power during the experiment. The power dependent PL plots [Figs. 4(a-c) and Figs. S4(a-c)] reveal that the emission intensity increases up to a maximum value and decreases thereafter with further increment of laser power for ML $WS_2$ on each substrate. This downfall in PL intensity is followed by a red-shift-blue-shift crossover (RBC) for excitonic resonance peak in case of dielectric substrates ($Al_2O_3$ and $SiO_2$/Si). PL spectra with varying excitation power have been deconvoluted to understand the peak shift and spectral weight variation of each quasiparticle with carrier density for all substrates [see Fig. S5]. In case of ML $WS_2$ on $Al_2O_3$, a sudden fall in the integrated intensity ($I_{PL}$) of exciton is noticed at $P_d \sim$ 8.55 $MW/cm^2$, followed by the RBC of resonance peak. For the $SiO_2$/Si substrate, the RBC phenomena and fall of intensity are perceived at $\sim$ 9.73 $MW/cm^2$ laser power density. This anomalous behavior in case of two different dielectric substrates is advantageous for achieving a high carrier density via optical doping in ML $WS_2$, followed by Mott insulator to metallic transition. Notably, for the ML $WS_2$ on conducting surface (Au/$SiO_2$/Si), the fall in $I_{PL}$ of excitonic emission happens at $P_d \sim$ 2.0 $MW/cm^2$ [Fig. S4], however the resonance peak position gets only red-shifted without any RBC effect, throughout our experimental laser power density range. This is attributed to the channeling of optically excited charge carriers of $WS_2$ through metallic interface and consequently failing to achieve the Mott carrier density threshold.

Furthermore, the variation of resonance peak position and integrated intensity of A-exciton around critical Mott transition point is presented as a function of laser power density induced carrier density ($c_d$) inside ML $WS_2$ for all three samples [see Figs. 4(d-f)], where data are presented above 3 $MW/cm^2$ power density. In this context, the excitation carrier density has been estimated from the power density of the laser, measured during the experiment. The laser spot is tightly focused on the sample surface to transfer maximum power to the system and the excitation carrier density has been estimated using the equation [30],

$$n_0 = \frac{Power\ density \times \tau o \times (I/I_0) \times A}{E} \qquad (2)$$

Where, $\tau_0$ is the life time of excitonic quasiparticles in ML TMD materials (~1 ps), A is the absorbance of suspended ML $WS_2$ at 2.33 eV (532 nm) excitation, which is around 19% [11]. Furthermore, the interference from the substrate affecting the excitation intensity has been estimated using,

$$\left(\frac{I}{I_0}\right) = \frac{4}{(1 + r.i.)^2} \qquad (3)$$

Where, $r.i.$ is the refractive index of the substrate material at 532 nm laser excitation. Accordingly, we have extracted the excitation carrier density from incident laser power density for all three substrates. Thereafter, the RBC of excitonic peak position is noticed at an elevated $c_d \sim 3.267 \times 10^{12} cm^{-2}$ ($2.267 \times 10^{12} cm^{-2}$) for $P_d \sim 9.73\ MW/cm^2$ ($8.55\ MW/cm^2$) in case of $WS_2$ on $SiO_2/Si$ ($Al_2O_3$) substrate. Here, Region I (green shaded region) is described as lower carrier density regime, where excitons exhibit insulating gas phase. On the contrary, excitons transformed into conducting electron-hole-plasma phase in Region-II (yellow shaded region), which is denoted as the higher carrier density regime. The appearance of RBC and abrupt fall of $I_{PL}$, reconfirm the absence of bound excitons and creation of separate charged quasiparticles beyond the Mott-point, in case of ML $WS_2$ on $Al_2O_3$ & $SiO_2$ substrates. For a comparative discussion, we have also tabulated all the extracted parameters regarding the Mott transition in Table II. On the contrary, for metallic Au substrate [see Fig. 4(f)], the excitonic peak gets only red-shifted from the beginning with increasing $P_d$ (or $c_d$) and RBC as well as the Mott-transition is not observed even at a high $P_d$ due to the direct charge transfer from ML $WS_2$ to Au layer. For metallic Au substrate, $I_{PL}$ [see Fig. 4(c)] increases rapidly in the beginning and then starts to decrease linearly with increasing $P_d$ due to lower density of exciton formation in $WS_2$.

The occurrence of RBC due to the formation of electron-hole-plasma (EHP) beyond Mott density can be analyzed with an analogy of attraction-repulsion crossover of exciton-exciton interactions in 2D systems [36] (Detailed discussion in APPENDIX C). In our case, the energy is shifted due to excitonic interactions for entire optical excitation range and can be modelled using two power laws, analogous with the Lennard-Jones potential between atoms, which can be written as,

$$\Delta E = \varepsilon \left[ \left( \frac{r_0}{r_s} \right)^{k1} - \left( \frac{r_0}{r_s} \right)^{k2} \right] \qquad (4)$$

Here, $\varepsilon$, $r_0$, k1, k2 are the fitting parameters, where, $r_s$ is the average distance between excitons, $r_0$ is the distance, where energy shift ($\Delta E$) becomes zero and $\varepsilon$ represents the $\Delta E$ value at Mott transition point, where attraction repulsion crossover between excitons takes place. We have fitted our experimental data with equation 4 [see Figs. 5(a-b)] for the dielectric substrates ($Al_2O_3$ & $SiO_2$). The extracted parameters are, $\varepsilon = 39.24\ meV$, $r_0 = 3.011\ nm$, $k1 = 8.25$ and $k2 = 1.52$, for $Al_2O_3$ and $\varepsilon = 54.25\ meV$, $r_0 = 2.686\ nm$, $k1 = 3.69$ and $k2 = 0.71$, for $SiO_2$, which are in well agreement with the previously reported results on CVD grown $MLWS_2$ on sapphire substrate [36].

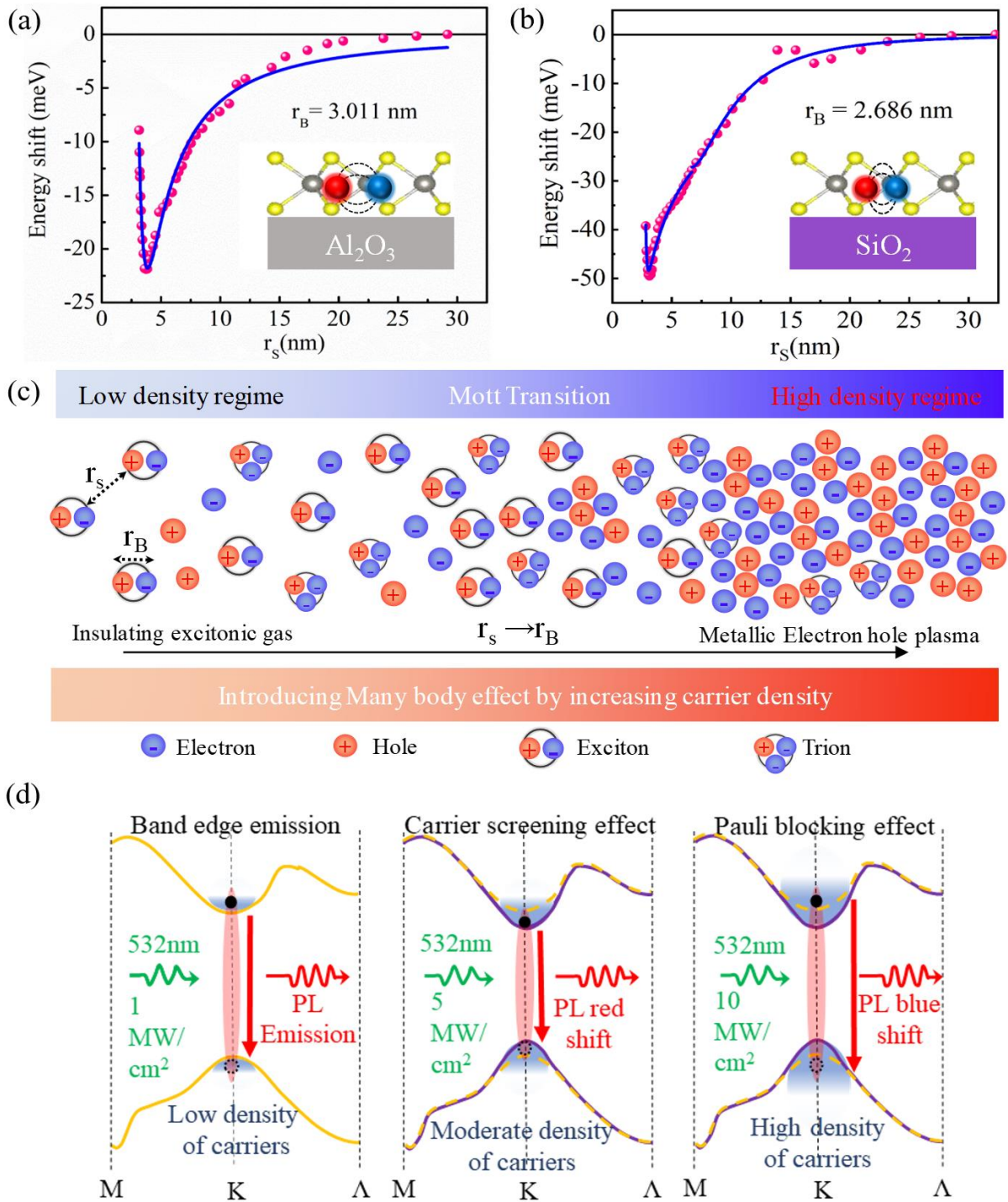

FIG.5. Exciton energy shift ($\Delta E$) variation with respect to average distance between excitons ($r_s$) for (a) $Al_2O_3$ and (b) $SiO_2$/Si substrates. The curves are fitted with Lennard–Jones potential and the excitonic Bohr radii are extracted for each substrate. In the inset of two figures, the schematic representation of exciton Bohr radius variation with dielectric environment alteration are illustrated. (c) Schematic of the increasing exciton density induced many-body effect in TMD materials leading to Mott transition in WS$_2$. (d) Schematic illustration of PL red-shift-blue-shift crossover due to interplay between Band renormalization and Pauli blocking effect of carriers.

The exciton Bohr radius ($r_B$) can be extracted from inter-excitonic distance around Mott point and usually, inter-excitonic distance inside atomically thin materials varies with carrier density as r$_s$= (1/πn)$^{1/2}$ [30]. With increasing photo-excitation, the excessive population of carriers in a specific area reduces inter-excitons separation in the high-density regime. Essentially, at a carrier density of $3.267 \times 10^{12} cm^{-2}$, the inter-excitonic distances are reduced to ∼ 3.12 nm, which is around the exciton Bohr radius of WS$_2$. At this extremum, excitons loss their identity and failed to be distinguished as e-h pairs. As a consequence, the whole exciton population starts to behave as free carriers, making resultant electron-hole-plasma (EHP) state by losing their binding energies to zero [36,52]. The obtained Bohr radius values from Lennard-Jones potential fittings reveal $r_B$ of ML WS$_2$ as $r_B = 2.686\ nm$ ($3.011\ nm$) for SiO$_2$ (Al$_2$O$_3$) substrates. In case of Au substrate, we are unable to calculate the Bohr radius of WS$_2$, as the Mott point is absent here. Now, the variation of Mott points as well as the Bohr radius of 2D ML WS$_2$ has a correlation with the substrate screening effect. The 2D dielectric constant ($\varepsilon_q^{2d}$) considering the surrounding medium can be expressed as,

$$\varepsilon_q^{2d} = \frac{k_a + k_b}{2} + qd\frac{k_{2D} - 1}{2} \qquad (5)$$

Where, $k_a$, $k_b$ are the dielectric constant of upper and lower dielectric materials and $k_{2D}$ is the dielectric constant of ML WS$_2$ [53]. Using the 2D model, we can evaluate the screened dielectric constant, which is more for high-k dielectric substrates. This in turn lowers the binding energy of excitons ($E_b^X$) and increases the excitonic Bohr radius defined by $r_B = \varepsilon_q^{2d} \hbar^2 / \mu e^2$, where $\mu = (m_e m_h / m_e + m_h)$ is the reduced effective mass of exciton. From the Lennard-Jones potential fit, we have extracted $r_B$ for SiO$_2$ to be lower than that for Al$_2$O$_3$, attributing to the substrate screening effect, which has a great impact on deciding the Mott-transition point with the variation of dielectric substrate. The dielectric screening induced exciton Bohr radius distinction is illustrated schematically in the bottom right inset of Figs. 5(a-b).

In a nutshell, the RBC and Mott transition phenomena of excitonic resonance energy with increasing P$_d$ (or c$_d$) are the result of different many-body effects acting synergistically on the excitonic quasiparticles in ML WS$_2$ as depicted schematically in Fig. 5(c-d). At a lower P$_d$, the screening of Coulomb repulsion among similar charges with increase in c$_d$ leads to decrease in the electronic band-gap, which is generally referred as band-gap renormalization ($\Delta E_g^{BGR}$) [27–29]. On the contrary, photogenerated excess carriers can screen the Coulomb attraction between electrons and holes, promoting the reduction of exciton binding energy ($\Delta E_b^{CP}$). Therefore, the position of excitonic transition with increasing carrier population is a consequence of the competitive behavior between the band-gap reduction ($\Delta E_g$) causing a red-shift and the binding energy reduction ($\Delta E_b$) causing a blue-shift in PL resonance energy. It is important to note here that $E_g$ decreases faster than $E_b$, so excitonic peak undergoes into a red-shift initially with increasing laser power [28,54]. Further increment of laser power leads to high carrier density regime, where the number of charge carriers increases to a higher value and due to strong interaction between them Mott-transition can occur at n > 2×10$^{12}$ cm$^{-2}$ [29]. In the Mott-transition regime, another phenomenon comes into play, namely Pauli blocking of carriers, which typically attributes to the blue-shift of exciton resonance beyond Mott point [55]. With further increase in $P_d$, the states near the conduction band edge start to fill up. This phase-space filling effect (or Moss-Brustein effect) [29] is governed by the Pauli exclusion principle due to the fermionic nature of electrons and holes, thus eventually pushes the free carriers to occupy higher energy states of the conduction band [56]. As a consequence, an apparent band-gap increase phenomenon, also referred to as Pauli blocking effect $(\Delta E_g^{PB})$ with reduced exciton oscillator strength, and corresponding decrement in exciton binding energy $(\Delta E_b^{PSF})$ is observed. Therefore, the pivotal expressions for realizing the RBC point can be written as,

$$\Delta E_g = \Delta E_g^{BGR} + \Delta E_g^{PB} \qquad (6)$$

$$\Delta E_b = \Delta E_b^{CP} + \Delta E_b^{PSF} \qquad (7)$$

As a result, the interplay among all these diverse many-body effects controls the optical transition rule in MLWS$_2$ with variable substrates, paving the way for added flexibility in designing novel optical devices for futuristic 2D materials based photonic systems.

## III.    SUMMARY AND CONCLUSION

In conclusion, a correlation between optical doping induced charge screening and substrate induced dielectric screening has been investigated in monolayer 2D WS$_2$ flakes, with the observed properties explained in the context of many-body phenomena that come into play at an elevated photo-excited carrier density ($\sim 10^{12}$ cm$^{-2}$ and above). A comparative study of steady-state emission and Raman spectroscopy for dielectric and metallic substrates, at a lower carrier density reveals substrate induced unintentional doping and dielectric screening on ML WS$_2$, as corroborated with our DFT results. Moreover, with increasing carrier density, a clear red-shift of excitonic resonance energy is observed for all three substrates, originating from carrier screening and long-range exciton-exciton attraction effects. However, in case of dielectric substrates, at around Mott-transition densities ($n_M \sim 3 \times 10^{12}$ cm$^{-2}$), a clear blue-shift is noticed, giving rise to novel RBC phenomena. The observed RBC phenomena on dielectric substrates are attributed to the combinatorial effect of Pauli blocking and short-range exciton-exciton repulsion, which is however absent for metallic Au substrate due to efficient charge transfer from WS$_2$. The observation could be explained by Lennard-Jones potential model for two dielectric substrates (SiO$_2$ & Al$_2$O$_3$), which also provides an estimate of the exciton Bohr radii for ML WS$_2$. The reported in-depth introspections on the synergistic behavior of the PL emission profile, RBC phenomena and Mott-transition points for a range of substrates are appealing in designing of numerous 2D TMD photonic devices operating at a wide excitation power regime (0.1–12.0 MW/cm$^2$).

## APPENDIX A: THEORETICAL ANALYSIS

The influence of different substrates on the electronic properties of WS$_2$ has been analyzed after investigating its interfaces with three different substrates, *viz.*, 1) Al$_2$O$_3$/WS$_2$, 2) SiO$_2$/WS$_2$ and 3) Au [111]/WS$_2$. The optical properties of WS$_2$ will be a closely related function of the interlayer coupling and therein generate mutual charge transfer within the constituent layers. The density functional investigations of these three WS$_2$ substrate interfaces were carried out by using the spin-polarized calculations with norm-conserving projector augmented wave (PAW) pseudopotentials, as implemented in the Vienna Ab initio Simulation Package (VASP). The exchange-correlation interactions were treated with the generalized gradient approximation (GGA) with Perdew-Burke-Ernzerhof (PBE) functions after incorporating the spin-orbit coupling (SOC). Interfacial dipolar van der Waals corrections were included via a semi-empirical dispersion potential to the DFT energy functional according to the Grimme DFT-D2 method [57]. The cut-off energy for the plane-wave expansion was set as 500 eV and a Monkhorst-Pack grid of $5 \times 5 \times 3$ was used for the Brillouin zone sampling for all calculations. The ionic positions and the lattice parameters were optimized by using the conjugate gradient algorithm until the Hellmann-Feynman force on each ion was less than 0.01 eV.

To reinvestigate substrate induced doping effects on ML WS$_2$, we calculated the electronic band structure of all the samples using density functional theory (DFT). According to our calculations, the up (red) and down (blue)-spin channel projections of W-5$d$ and both spin channel projections of S-3$p$ orbitals (green) are presented for WS$_2$ monolayer in Fig. 3(i). The bonding and anti-bonding manifolds are populated by S-3$p_x$ and 3$p_y$-hybridized W-5$d_{xz}$, 5$d_{yz}$, 5$d_{xy}$ and S-3$p_z$ hybridized W-5$d_{x^2-y^2}$ and W-5$d_{3z^2-1}$ states, respectively. The layer projections of Al$_2$O$_3$ and WS$_2$ (defined as the interface-1) are presented in Figs. 3(ii) and (iii), respectively with the color codes as defined in the respective figures. The interface, having the lowest interfacial lattice mismatch, has the least doping and thereby succeeds to generate a layer-decoupling of excitonic quasiparticles. Interface 2 (WS$_2$-SiO$_2$), exhibits an increase of the $n$-type doping in this system, as compared to the earlier one. Figs. 3(iv) and (v) represent the layer projections of the SiO$_2$ and WS$_2$ levels of the interface 2, respectively. Whereas the valence levels are mostly populated by the O-2$p$ and hybridized W-5$d$ and S-3$p$ states, the conduction band is contributed by the Si-3$s$ and 3$p$ states, indicating transfer of electrons from Si-3$s$ and 3$p$ to WS$_2$ via O-2$p$ levels. Interface 3 (WS2-Au), on the other hand, represents the largest amount of charge-transfer from the Au[111] surface to the WS$_2$ layer and as a result, the entire interface acquires a metallic nature. Figs. 3(vi) and (vii) illustrate the layer projections of Au[111] and WS$_2$ layers, indicating highly delocalized Au-5$d$ and 6s levels around $E_F$, transfering the charges to the S-3p levels of WS$_2$ and thereby leading to highly hybridized Au and S levels. On the other hand, the density of states (DOS) for these three interfaces are depicted in Fig S2. As will be evident from an overall comparison of these two figures, there are monotonic incremental shifts of the E$_F$ for these three interfaces towards the conduction band of the combined system, indicating a gradual increase of the *n*-type doping from the interface system 1 to 3, as corroborated with our Raman results. This increase of the *n*-type doping can be attributed to the increase of mutual charge transfer (electron) from the substrate to the ML WS$_2$. The extents of doping are seen to have an explicit relationship with the mismatch of lattice parameters of the constituent layers. The extent of interlayer charge transfer and thereby induced explicit *n*-type doping are more evident from the Fig. S2(b), where the converged charge-density distributions are presented. Gradual increase of charge transfer manifesting larger interfacial overlap of charge spheres from interface 1 to 3 is obvious from the Fig. S2b (i-iii).

**APPENDIX B: DIELECTRIC SCREENING INDUCED BAND RENORMALIZATION**

According to Fig. 2(b), a clear PL peak shift is noticed which is a direct evidence of dielectric screening induced band renormalization. At a relatively low excitation power (neglecting Pauli blocking effect), the electronic band-gap can be written as,

$$E_g^X = E_g^{SPX} + E_g^{BGRX0}/\left(k_{eff}\right)^{\beta} \tag{D1}$$

Where, $E_g^{SPX}$ is the electronic band-gap from single particle model, $E_g^{BGRX0}$ is the band-gap renormalisation term in vacuum, $\beta$ is another scaling factor and $X$ represents quasiparticles generated inside WS$_2$. Hence, for a higher-k dielectric (Al$_2$O$_3$), the effect of band renormalization and hence $E_g^X$ will be lesser than SiO$_2$ with a lower- dielectric permittivity. On the other hand, the binding energy ($E_b$) of an excitonic quasiparticle reflects the Coulomb potential between the charges, which is screened further by the dielectric environmental [58]. The variation of binding energy of quasiparticles with effective dielectric constant ($k_{eff}$) of the environment can be described by,

$$E_b^X = E_b^{X0}/(k_{eff})^\alpha \tag{D2}$$

Where, $E_b^{X0}$ is binding energy in vacuum (without substrate), $k_{eff} = (k_1 + k_2)/2$ is the effective dielectric constant. Here, $k_1$ & $k_2$ are the dielectric constants of the top & bottom dielectric layers of the WS$_2$ flake and $\alpha$ is the scaling factor. So, the binding energy of quasiparticles ($E_b^X$) will reduce with increasing dielectric constant of the environment causing a blue shift of energy. The optical band-gap or quasiparticle resonance energy can be described as,

$$PL_X = E_g^X - E_b^X \tag{D3}$$

$$PL_X = E_g^{SPX} + E_g^{BGRX0}/(k_{eff})^\beta - E_b^{X0}/(k_{eff})^\alpha \tag{D4}$$

As $E_b^X$ is strongly affected via dielectric screening than $E_g^X$, a blue shift of the PL peak is observed for higher-k dielectric Al$_2$O$_3$, as compared to SiO$_2$.

### APPENDIX C: MODELLING OF RBC PHENOMENA

Exciton-exciton interactions are modulated by generating substantial carrier density inside ML WS$_2$, which initiates the transition from an insulating excitonic state to the metallic electron-hole plasma state. Similar to the hydrogen atom, the attractive exciton-exciton interaction matches with the long-range Lennard-Jones interaction between atoms. Mutual attraction between any two excitons reduces the energy required to form an additional exciton by an amount of the negative interaction potential energy, which results in the red-shift of PL resonance peak. Using the same analogy, the origin of excitonic blue-shift can be interpreted for higher carrier densities at a higher laser power. Large fraction of charge carriers creates high density of bound excitons via reducing the inter-excitonic distances, leading to the stronger exciton-exciton repulsion due to overlapping electronic orbitals. This process leads to positive inter-exciton potential energy, which is the plausible reason behind the excitonic blue-shift. Considering excitons as rigid bodies, the critical densities at which the excitons are closely packed, can be determined as, $n_e \pi r_s^2 \sim 1$, where, $n_e$ is the density of the excitons and $r_s$ is the average distance between excitons [30]. Around the Mott-transition point, the exciton density becomes so high such that the inter excitonic distance reduces to excitonic Bohr radius ($r_B$) of the material.

Table I. Substrate induced charge doping

| Substrate | $I_A$ | $I_{A-}$ | $I_A : I_{A-}$ | Doping | $n_e$ ($10^{13}$ cm$^{-2}$) |
|-----------|-------|----------|----------------|--------|------------------------------|
| SiO$_2$ | 3874.57 | 5989.36 | 1 : 1.54 | Highly n-doped | 3.62 |
| Al$_2$O$_3$ | 774.80 | 478.80 | 1 : 0.62 | n-doped | 1.54 |
| Au | 6028.94 | 2538.43 | 1 : 0.42 | Lightly n-doped | 0.98 |

Table II. Substrate dependent Mott transition point and Bohr radius

| Substrate | Mott point (MW.cm$^{-2}$) / and carrier density (cm$^{-2}$) | Red shift amount (meV/ MW.cm$^{-2}$) | Blue shift amount (meV/ MW.cm$^{-2}$) | Bohr radius (nm) |
|-----------|------------------------------------------------------------|--------------------------------------|---------------------------------------|-------------------|
| SiO$_2$/Si | 9.73/ $3.267 \times 10^{12}$ | 4.82 | 4.51 | 2.686 |
| Al$_2$O$_3$ | 8.55/ $2.267 \times 10^{12}$ | 2.14 | 3.76 | 3.011 |

ASSOCIATED CONTENT

The Supplemental Material consists of Figs. S1−S6, detailed optical analysis for power dependent PL and dielectric screening effect inside $WS_2$.

AUTHOR INFORMATION

**Corresponding Author**

Samit K. Ray - Department of Physics, Indian Institute of Technology Kharagpur, Kharagpur - 721302, India; Email: physkr@phy.iitkgp.ac.in.

**Authors**

Shreyasi Das - School of Nano Science and Technology, Indian Institute of Technology Kharagpur, Kharagpur-721302, India.

Rup K. Chowdhury - Department of Physics, Indian Institute of Technology Kharagpur, Kharagpur -721302, India.

Present address - Department of Ultrafast Optics and Nanophotonics, IPCMS, CNRS, Strasbourg-67034, France.

Debjani Karmakar - Technical Physics & Prototype Engineering Division, Bhabha Atomic Research Center, Mumbai-400085, India.

Soumen Das - School of Medical Science and Technology, Indian Institute of Technology Kharagpur, Kharagpur-721302, India.

**Author Contributions**

The manuscript was written through contributions of all authors. All authors have given approval to the final version of the manuscript.

**Notes**

The authors declare no competing financial interest.